\documentclass[a4paper,prb,twocolumn,superscriptaddress]{revtex4}

\usepackage{bm}
\usepackage{color}
\usepackage{graphicx}
\usepackage{stmaryrd}

\begin{document}

\title{Ion association in low-polarity solvents: comparisons between 
theory, simulation, and experiment}

\author{Chantal Valeriani}

\affiliation{SUPA, School of Physics and Astronomy, The University of 
Edinburgh, Mayfield Road, Edinburgh EH9 3JZ, UK}

\affiliation{Soft Condensed Matter, Debye Institute for NanoMaterials 
Science, Utrecht University, Princetonplein 5, 3584 CC Utrecht, The 
Netherlands}

\author{Philip J. Camp}

\email[Corresponding author:~~]{philip.camp@ed.ac.uk}

\affiliation{School of Chemistry, The University of Edinburgh, West 
Mains Road, Edinburgh EH9 3JJ, UK}

\author{Jos W. Zwanikken}

\affiliation{Institute for Theoretical Physics, Utrecht University, 
Leuvenlaan 4, 3584 CE Utrecht, The Netherlands}

\affiliation{Department of Material Science and Engineering, 
Northwestern University, 2220 Campus Drive, Evanston, Illinois 
60208-3108, US}

\author{Ren{\'e} van Roij}

\affiliation{Institute for Theoretical Physics, Utrecht University, 
Leuvenlaan 4, 3584 CE Utrecht, The Netherlands}

\author{Marjolein Dijkstra}

\affiliation{Soft Condensed Matter, Debye Institute for NanoMaterials 
Science, Utrecht University, Princetonplein 5, 3584 CC Utrecht, The 
Netherlands}

\begin{abstract} The association of ions in electrolyte solutions at 
very low concentration and low temperature is studied using computer 
simulations and quasi-chemical ion-pairing theory. The specific case of 
the restricted primitive model (charged hard spheres) is considered. 
Specialised simulation techniques are employed that lead to efficient 
sampling of the arrangements and distributions of clusters and free 
ions, even at conditions corresponding to nanomolar solutions of simple 
salts in solvents with dielectric constants in the range $5$-$10$, as 
used in recent experimental work on charged-colloid suspensions. A 
direct comparison is effected between theory and simulation using a 
variety of clustering criteria and theoretical approximations. It is 
shown that conventional distance-based cluster criteria can give 
erroneous results. A reliable set of theoretical and simulation 
estimators for the degree of association is proposed. The ion-pairing 
theory is then compared to experimental results for salt solutions in 
low-polarity solvents. The agreement is excellent, and on this basis 
some calculations are made for the screening lengths which will figure 
in the treatment of colloid-colloid interactions in such solutions. The 
accord with available experimental results is complete. \end{abstract}

\date{\today}

\maketitle

\section{Introduction} \label{sec:introduction}

The control of charged-colloid suspensions with added salt is a linchpin 
of soft condensed matter science. The physical principles that govern 
the net interactions between like-charged colloids in aqueous 
electrolyte solutions were laid down more than half a century ago. 
\cite{Verwey:1948/a} Central to charge stabilization are the formation 
of the electrical double layer and the phenomenon of screening over 
distances comparable to the Debye length. The classic 
Derjaguin-Landau-Verwey-Overbeek and Debye-H{\"u}ckel (DH) theories 
apply well to high-polarity solvents, such as water, where the 
electrostatic interactions between salt ions and counterions are 
strongly screened dielectrically. In particular, under normal conditions 
and with simple salts, cation-anion pairing is rather insignificant and 
complete ion dissociation can be assumed when describing the screening 
effect.

Charged colloidal suspensions in low dielectric constant solvents are 
now of experimental interest. 
\cite{Yethiraj:2003/a,Royall:2003/a,Leunissen:2005/a,% 
Campbell:2005/a,Shevchenko:2006/a,Lu:2006/a,Leunissen:2007/a,% 
Leunissen:2007/b,Leunissen:2007/c,Zwanikken:2007/a} Such systems involve 
whole new regimes of low ion concentrations $\sim\mbox{nM}$, and strong 
electrostatic interactions between ions (added salt and counterions). 
The situation with regard to screening now changes because the reduction 
in dielectric screening compared to that in water can only enhance 
cation-anion association and promote the formation of a significant 
number of so-called Bjerrum pairs. This leads to a reduction in free 
ions and a concomitant increase in the screening length. The effects of 
ion pairing on the screening of colloidal interactions have been 
explored. \cite{Allahyarov:2007/a,Allahyarov:2008/a,Zwanikken:2009/a}

Of course, ion pairing is not a new phenomenon, and its effects on the 
thermodynamics and dynamical properties can be highly pronounced. 
\cite{Robinson:1959/a} Already in 1926, Bjerrum described his eponymous 
pairs within a quasi-chemical ion-pairing equilibrium, and suggested 
removing them from the effective free-ion concentration when performing 
DH-like calculations on electrolyte solutions. \cite{Bjerrum:1926/a} 
Bjerrum's theory has been thoroughly tested against experimental data 
for solutions with moderately low salt concentrations of $c_{\rm s} \ge 
10^{-5}~\mbox{M}$ in solvents with dielectric constants in the range $2 
\leq \epsilon \leq 80$. \cite{Robinson:1959/a} One of the most dramatic 
manifestations of ion pairing is in the phase separation of ionic 
fluids, \cite{Pitzer:1984/a,Pitzer:1990/a} where the low-concentration 
`vapour' phase has such a high degree of ion association that the 
conventional DH theory has to be extended to include ion-dipole and 
dipole-dipole interactions in order to give a good account of accurate 
simulation data for the coexistence envelope. 
\cite{Fisher:1993/a,Levin:1996/a}

From a computational perspective, the new experimental regimes of very 
low concentration and strong electrostatic interactions present some 
serious challenges. Molecular dynamics simulations of salts at nanomolar 
concentrations have already fallen foul of sampling problems. 
\cite{Allahyarov:2007/a,Allahyarov:2008/a} Recently, the current authors 
put forward a protocol for performing Monte Carlo (MC) simulations in 
the canonical ensemble, with novel particle moves that allow efficient 
equilibration at the extreme conditions referred to above. 
\cite{PJC:2010/b} This opens up the opportunity to explore the true 
degree of association in very low concentration electrolyte solutions 
made up with low-polarity solvents. For the purposes of this exploratory 
study, attention is focused on the restricted primitive model (RPM) of 
ionic fluids. The RPM is an electroneutral mixture of $N/2$ positively 
and $N/2$ negatively charged hard spheres of equal diameter $\sigma$ and 
charges $\pm q$ immersed in a dielectric continuum with dielectric 
constant $\epsilon$ and volume $V$ at temperature $T$, with an overall 
ion concentration $\rho=N/V$. The interaction pair potential between 
ions $i$ and $j$ is
\begin{equation} 
u(r_{ij}) = 
 \left\{ 
  \begin{array}{ll} 
   \infty                                  & r_{ij} <    \sigma \\ 
   \displaystyle\frac{q_{i}q_{j}}{Dr_{ij}} & r_{ij} \geq \sigma 
  \end{array} 
 \right. 
\end{equation} 
where $r_{ij}$ is the pair separation, $q_{i}$ is the charge on ion $i$, 
and $D=4\pi\epsilon_{0}\epsilon$ where $\epsilon_{0}$ is the dielectric 
permittivity of vacuum. The overall ion concentration and temperature 
are given in reduced units by $\rho^{*}=\rho\sigma^{3}$ and 
$T^{*}=k_{\rm B}TD\sigma/q^{2}$, respectively. The Bjerrum length is the 
distance at which the attractive cation-anion potential is equal to 
$-k_{\rm B}T$, and is given by $\lambda_{\rm B}=\sigma/T^{*}$. The phase 
behaviour of the RPM is now well known; \cite{Romero:2002/a,Vega:2003/a} 
the vapour-liquid critical parameters are $T_{\rm c}^{*} \simeq 0.05$ 
and $\rho_{\rm c}^{*} \simeq 0.08$. \cite{Caillol:2002/a,Luijten:2002/b} 
Simulations confirm that the degree of ion association in the vapour 
phase just below $\rho_{\rm c}^{*}$ and $T_{\rm c}^{*} \simeq 0.05$ is 
significant. 
\cite{Valleau:1980/a,Gillan:1983/a,Caillol:1995/a,Bresme:1995/a,% 
PJC:1999/a,PJC:1999/c} Indeed, the coexistence properties of fused 
cation-anion pairs (charged hard dumbbells) are almost identical to 
those of the RPM. \cite{Shelley:1995/a,PJC:2003/d,PJC:2007/b}

In this work, the degree of ion association in the RPM at very low 
concentrations and low (near-critical) temperatures is investigated. 
Calculations are performed down to a reduced ion concentration of 
$\rho^{*}=10^{-10}$ and a reduced temperature of $T^{*}=0.04$; for a 
monovalent salt with ionic diameter $\sigma=4~\mbox{\AA}$ at room 
temperature, these values correspond to a salt concentration $c_{\rm s} 
\simeq 1~\mbox{nM}$ and a solvent dielectric constant $\epsilon \simeq 
5.6$. Using specialised MC simulations, results are obtained with which 
to test the quasi-chemical ion-pairing theory as proposed by Bjerrum. 
This involves using a novel simulation protocol recently proposed by us, 
\cite{PJC:2010/b} and a variety of methods for determining the degree of 
association. The ion-pairing theory is then tested against experimental 
data for the degree of ion association; to this end, recent work by 
Leunissen and co-workers has yielded results for salts at concentrations 
of around $10^{-7}~\mbox{M}$ in solvents with dielectric constants as 
low as about $5$. \cite{Leunissen:2007/c,Hollingsworth:2010/a} To the 
best of our knowledge, this is the first time that a quantitative 
comparison has been made between theory, simulation, and experiment at 
such extreme conditions. On the basis of this comparison, the effects of 
ion association on the screening of charged-colloid interactions under 
such conditions can be evaluated with some confidence.
	
This article is arranged as follows. The ion-pairing theory is presented 
in Section \ref{sec:theory}, and the simulation details are summarised 
in Section \ref{sec:simulations}. The RPM simulation results are given 
in Section \ref{sec:rpm_results}, and an analysis of experimental data 
is presented in Section \ref{sec:exp_results}. Section 
\ref{sec:conclusions} concludes the paper.

\section{Theory} \label{sec:theory}

To describe ion association, consider the quasi-chemical equilibrium
\begin{equation}
\mbox{cation-anion pair} 
\rightleftharpoons
\mbox{cation}+\mbox{anion}.
\end{equation}
In terms of the degree of association $\alpha$, the concentration of 
cation-anion pairs is $\rho_{\pm}=\alpha\rho/2$ and those of the cations 
and anions are $\rho_{+}=\rho_{-}=(1-\alpha)\rho/2$. Considering the 
mixture of cations, anions, and cation-anion pairs to be ideal, the 
chemical potentials are $\mu_{\pm}=k_{\rm 
B}T\ln{(\alpha\rho\Lambda_{+}^{3}\Lambda_{-}^{3}/2K)}$, $\mu_{+}=k_{\rm 
B}T\ln{[(1-\alpha)\rho\Lambda_{+}^{3}/2]}$, and $\mu_{-}=k_{\rm 
B}T\ln{[(1-\alpha)\rho\Lambda_{-}^{3}/2]}$, where $\Lambda_{+}$ and 
$\Lambda_{-}$ are the de Broglie thermal wavelengths of the cations and 
anions, respectively. Here $K$ is the configurational integral of a 
pair, which plays the role of an equilibrium constant:
\begin{equation}
K = 4\pi\int_{\sigma}^{r_{\rm c}} 
    r^{2}\exp{\left(\frac{\sigma}{rT^{*}}\right)}
    {\rm d}r.
\label{eqn:k}
\end{equation}
The choice of the cutoff radius $r_{\rm c}$ is to be discussed below. At 
equilibrium $\mu_{\pm}=\mu_{+}+\mu_{-}$, which leads to
\begin{equation}
\frac{\alpha}{(1-\alpha)^{2}} = \frac{K\rho}{2}.
\end{equation}
Solving for $\alpha$ yields
\begin{equation}
\alpha = 1
       - \frac{1}{K\rho}
         \left(
          \sqrt{1+2K\rho} - 1
         \right).
\label{eqn:alpha}
\end{equation}
Considering the phase diagram of the RPM in the 
concentration-temperature plane, a sensible dividing line between 
`associated' and `dissociated' regimes is the locus of points defined by 
$\alpha=\frac{1}{2}$, or alternatively
\begin{equation}
K\rho = 4.
\label{eqn:bjlocus}
\end{equation}
All that remains now is to determine the equilibrium constant $K$. The 
primary problem is that the integral in Eq.~(\ref{eqn:k}) does not 
converge for $r_{\rm c} \rightarrow \infty$, and so in the conventional 
treatment, an appropriate finite upper limit for the integral needs to 
be identified. One choice for $r_{\rm c}$ is the Bjerrum length 
$\lambda_{\rm B}$, on the basis that the separation between ions in a 
pair should be such that the interaction energy is greater in magnitude 
than $k_{\rm B}T$. An alternative, and more conventional, choice is to 
set $r_{\rm c}=\lambda_{\rm B}/2$ corresponding to the minimum of the 
integrand $r^{2}\exp{(\lambda_{\rm B}/r)}$ in Eq.~(\ref{eqn:k}). It has 
long been recognised, however, that the precise choice of $r_{\rm c}$ is 
unimportant (see section 925 of Ref.~\onlinecite{Fowler:1956/a}), at 
least at low temperatures; this will be emphasised in the results of the 
current work. An approximate closed-form expression for $K$ valid at low 
$T^{*}$ can be obtained by noting that in a cation-anion pair, the 
separation $r$ should not be much more than $\sigma$. Writing 
$r=\sigma+\delta r$ leads to the limiting behaviour $\sigma/r \approx 1 
- \delta r / \sigma = 2 - r/\sigma$. Substituting this in to 
Eq.~(\ref{eqn:k}) and performing the integral with $r_{\rm c}=\infty$ 
yields
\begin{equation}
K \approx 4\pi\sigma^{3}e^{1/T^{*}}
          \left[ 
           T^{*} + 2(T^{*})^{2} + 2(T^{*})^{3}
          \right].
\label{eqn:kasymp}
\end{equation}
Equation (\ref{eqn:kasymp}) is possibly the most simple low-temperature 
result, and was inspired by a similar approximation for the two-particle 
partition function of dipolar hard spheres presented by Jordan; 
\cite{Jordan:1973/a} the range of validity is limited by an unphysical 
minimum in $K$ at $T^{*} \simeq 0.54$. Levin and Fisher have summarised 
several more accurate closed-form expressions. \cite{Levin:1996/a} 
Finally, it is acknowledged that Ebeling's alternative expression for 
$K$, \cite{Ebeling:1968/a} which reproduces the correct equation of 
state for the RPM up to terms of order $\rho^{5/2}$ and is therefore a 
more rigorous choice, \cite{Levin:1996/a} gives essentially identical 
results to the Bjerrum-length prescriptions employed here.

\section{Simulation methods} \label{sec:simulations}

Conventional MC simulations of associating fluids at very low 
concentrations can fail due to insufficient sampling of the most 
significant arrangements and spatial distributions of clusters. 
\cite{Orkoulas:1994/a} On the one hand, during a typical length run 
using single-particle moves, isolated particles in very dilute systems 
may never come within sufficient proximity of other particles to 
associate. On the other hand, particles already within clusters may not 
be able to detach due to it being a rare event. In an effort to 
eliminate these problems, the authors recently proposed an efficient MC 
protocol for simulating the RPM at very low concentrations and low 
temperatures where ion association is expected to be significant. 
\cite{PJC:2010/b} The simulations are conducted within the canonical 
($NVT$) ensemble using a cubic box of side $L=V^{1/3}$ with periodic 
boundary conditions. The long-range coulombic interactions are handled 
using the Ewald summation with conducting boundary conditions. 
\cite{Allen:1987/a} Various types of MC moves are attempted: normal 
single-particle moves with displacements chosen randomly from either a 
narrow interval (with a width adjusted to give an acceptance rate of 
40\%) or a broad interval (spanning the range $-L/2$ to $L/2$); cluster 
moves with displacements chosen randomly from narrow and broad 
intervals, as before; and `cluster formation/breakage' (CFB) moves, each 
of which involves moving a second ion within a sphere of radius $\Delta$ 
centered on a randomly chosen first ion. This last move offers 
possibilities for bringing together two randomly selected isolated ions 
in to association, and for prising two clustered ions apart. Full 
details of the simulation protocol are reported in 
Ref.~\onlinecite{PJC:2010/b}. The main control parameters are the radius 
$\Delta$, and the various proportions of single-particle and cluster 
moves, small and large displacements, and CFB moves. On the basis of 
earlier work, \cite{PJC:2010/b} the present simulations are performed 
with 70\% small single-particle displacements, 10\% large 
single-particle displacements, 5\% small cluster displacements, 5\% 
large cluster displacements, and 10\% CFB moves. The CFB radius was set 
to $\Delta=L/4$ in all cases. These parameters were shown in 
Ref.~\onlinecite{PJC:2010/b} to give rapid convergence to the apparent 
equilibrium state. In all cases, the system is made up of $N=256$ ions, 
and run lengths consist of $10^{5}$-$10^{6}$ MC moves per ion, depending 
on density and temperature.

Conventionally, clusters in fluids are identified using some kind of 
pairwise distance \cite{Gillan:1983/a} or energy-based criterion; the 
latter are useful for anisotropic potentials, where not only the 
distance but also the orientation have to be favorable for association 
to occur. In the present case, a distance-based criterion suffices; two 
particles are considered associated if their separation is less than 
some cutoff distance $r_{\rm c}$. In their comprehensive study of ion 
association in the vapour phase of the RPM near coexistence, Caillol and 
Weis showed that the cluster distribution is basically independent of 
criteria in the range $1.8\sigma \leq r_{\rm c} \leq 2.2\sigma$. 
\cite{Caillol:1995/b} Allahyarov {\it et al.} use $r_{\rm 
c}=\lambda_{\rm B}$ in Ref.~\onlinecite{Allahyarov:2007/a} and $r_{\rm 
c}=3\sigma$ in Ref.~\onlinecite{Allahyarov:2008/a}. In this work, three 
different distance criteria were employed: $r_{\rm c}=\lambda_{\rm B}$ 
and $r_{\rm c}=\lambda_{\rm B}/2$ are obvious candidates, for the 
reasons outlined in Section \ref{sec:theory}; and $r_{\rm c}=2\sigma$, 
in line with earlier studies. \cite{Caillol:1995/b} Using these 
criteria, $\alpha$ is the proportion of ions clustered with at least one 
other ion.

In addition, outlined here is a method of estimating the degree of 
association $\alpha$ from simulation data without having to specify a 
cluster criterion. Consider the nearest-neighbour cation-anion 
distribution function $p(r)$, reflecting the distance between an ion and 
its nearest neighbour {\em of opposite charge}. If a cation is 
dissociated, then the nearest-neighbour anion is remote (due to the low 
densities of interest here) and to a first approximation can be assumed 
completely uncorrelated with the cation. The probability of an anion 
being at a distance between $r$ and $r+{\rm d}r$ from the cation, and 
the remaining $N/2-1$ anions being at least as far away, is
\begin{eqnarray}
p_{\rm d}(r){\rm d}r 
&=&       \frac{N}{2}
          \times \frac{4\pi r^{2}{\rm d}r}{V}
          \times \left( 1 - \frac{4\pi r^{3}}{3V} \right)^{N/2-1}
          \nonumber \\
&\approx& 2\pi\rho r^{2} 
          \exp{\left(-\mbox{$\frac{2}{3}$}\pi\rho r^{3}\right)}{\rm d}r
\end{eqnarray}
where the subscript `d' denotes `dissociated'. Note that $p_{\rm d}(r)$ 
is normalised and shows a peak at $r_{0} = (1/\pi\rho)^{1/3}$. If the 
nearest-neighbour anion is associated with the cation, then the radial 
distribution function $g_{+-}(r)$ will be peaked near $r=\sigma$, 
signalling very strong, short-range correlations which are not amenable 
to an accurate theoretical treatment; the corresponding function for 
associated (`a') cations, $p_{\rm a}(r)$, is not easy to predict. For 
the RPM the arguments above apply in exactly the same way to anions. If 
the proportion of associated ions is $\alpha$, and that of dissociated 
ions is $(1-\alpha)$, then the total $p(r)$ will be given by
\begin{equation}
p(r) = \alpha p_{\rm a}(r) + (1-\alpha) p_{\rm d}(r).
\label{eqn:pr}
\end{equation}
This function can be obtained directly from simulations and, in 
principle, fitting Eq.~(\ref{eqn:pr}) to simulation results yields the 
degree of association without having to specify a cluster criterion. In 
practice, and without a reliable expression for $p_{\rm a}(r)$, 
$(1-\alpha)p_{\rm d}(r)$ is fitted to $p(r)$ over the range $r \geq 
r_{0}$, where $\alpha p_{\rm a}(r)$ makes no significant contribution:
\begin{equation}
p(r) \simeq (1-\alpha)
            \left[
             2\pi\rho r^{2} 
             \exp{\left(-\mbox{$\frac{2}{3}$}\pi\rho r^{3}\right)}
            \right]~~~~r>r_{0}.
\label{eqn:prd}
\end{equation}

\section{Results} \label{sec:results}

\subsection{Restricted primitive model} \label{sec:rpm_results}

\begin{figure}[tb]

\begin{tabular}{cc}
\includegraphics[width=4.25cm]{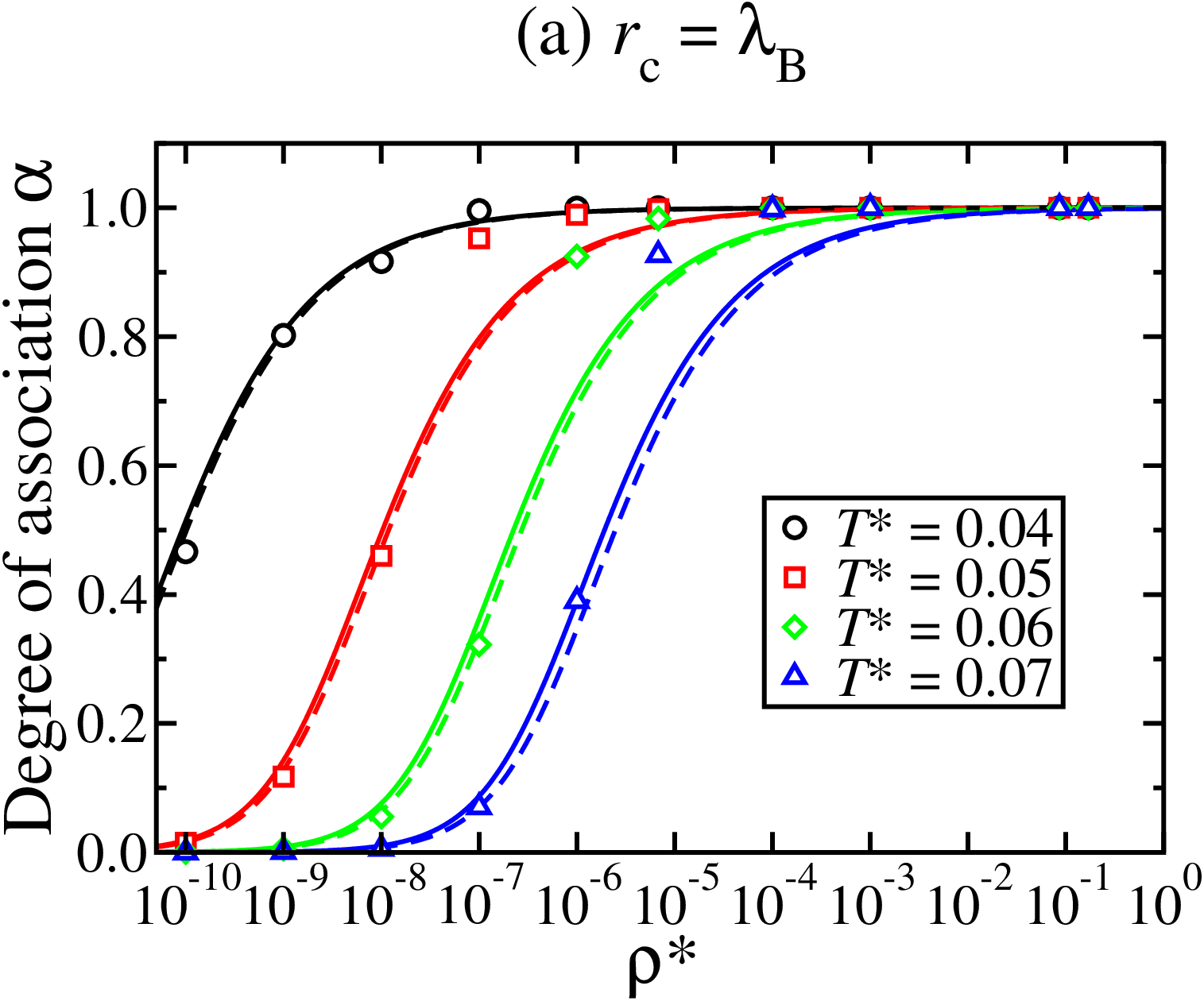} &
\includegraphics[width=4.25cm]{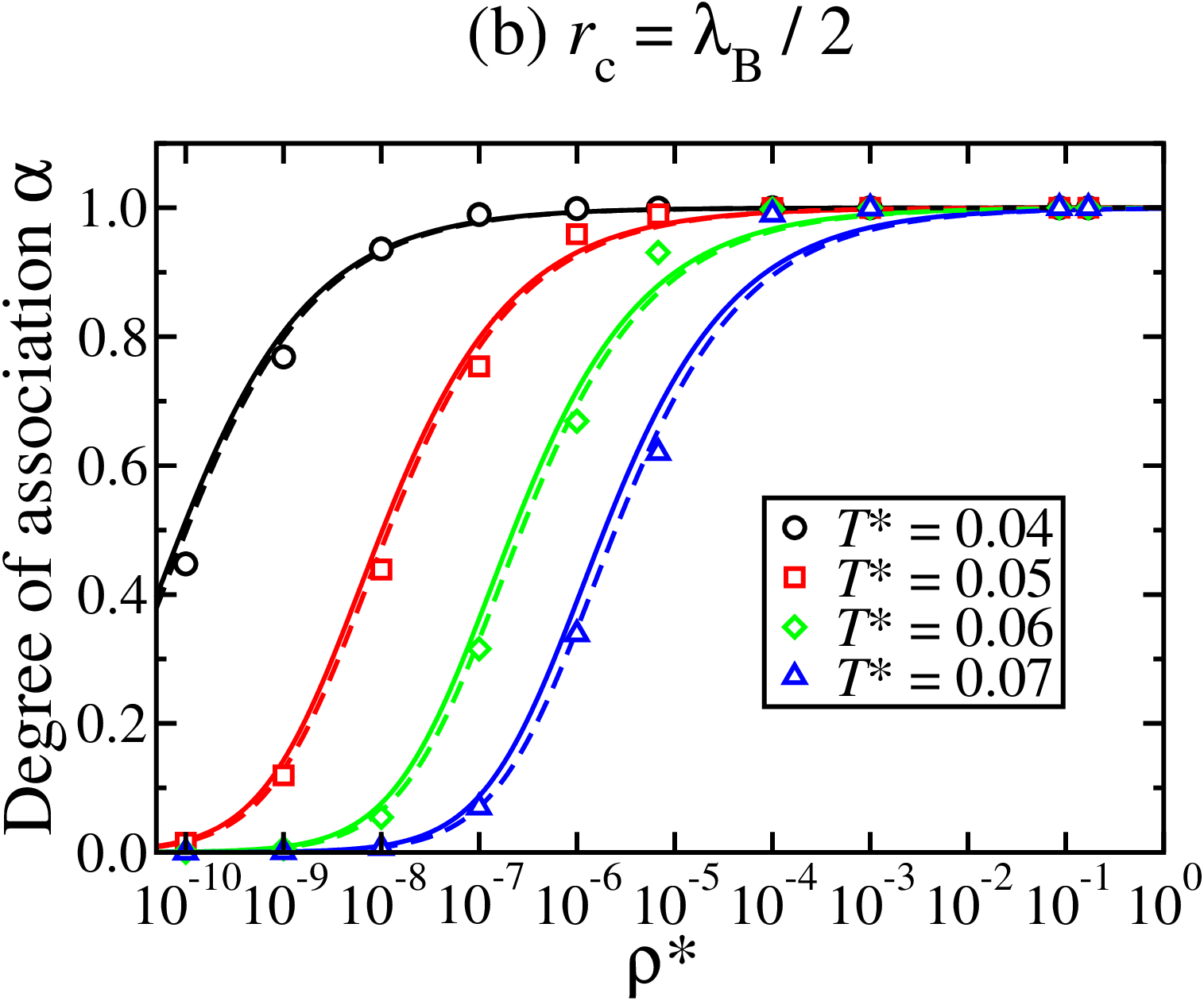} \\
\includegraphics[width=4.25cm]{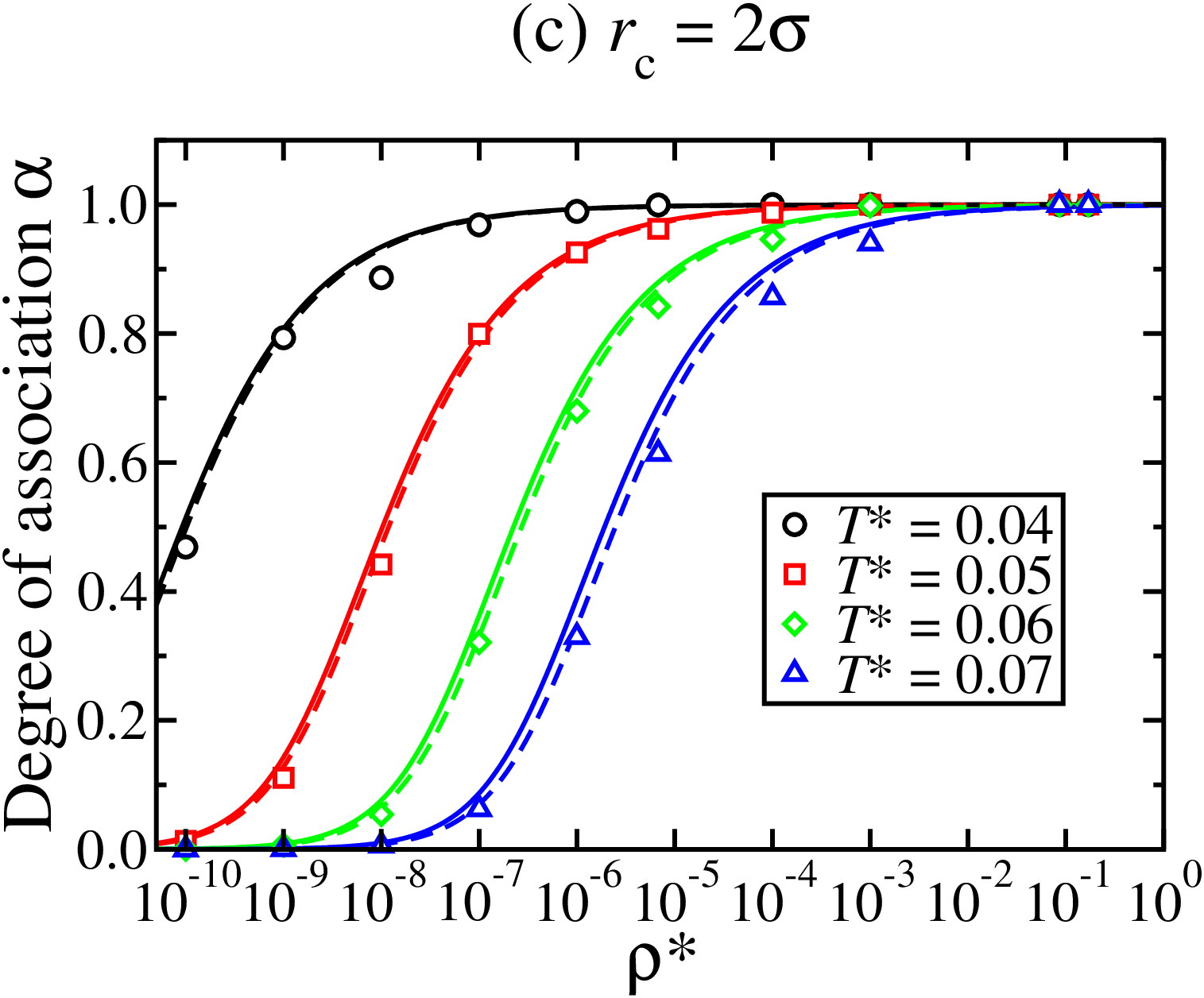} &
\includegraphics[width=4.25cm]{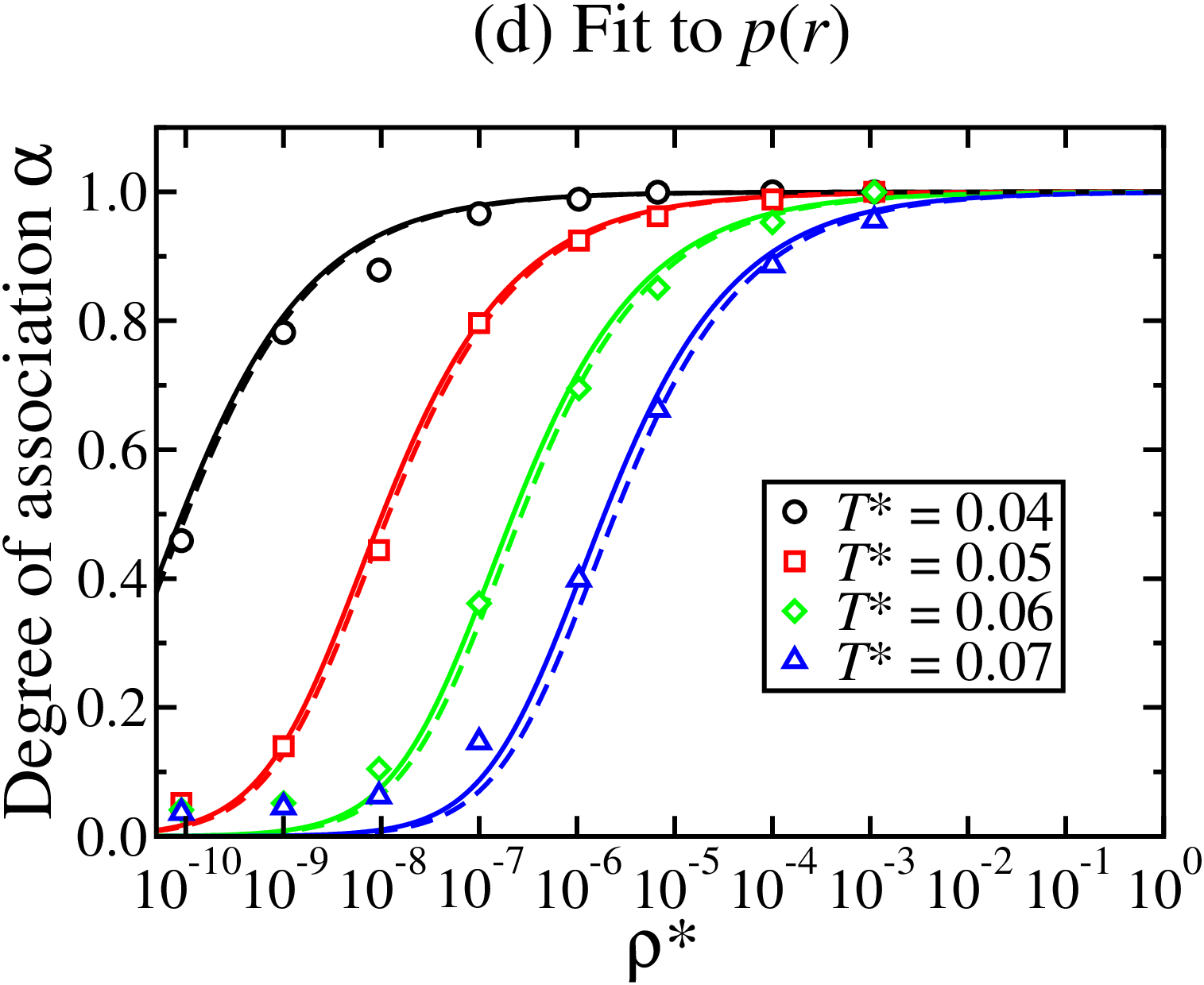} \\
\end{tabular}

\caption{\label{fig:alpha} The degree of association $\alpha$ along 
isotherms. The curves are the theoretical predictions of 
Eq.~(\ref{eqn:alpha}) using $K$ computed with a cut-off $r_{\rm 
c}=\lambda_{\rm B}/2$ (solid lines), and using the asymptotic result in 
Eq.~(\ref{eqn:kasymp}) for $K$ (dashed lines). The points are the 
simulation results computed using various criteria: (a) distance 
criterion with $r_{\rm c}=\lambda_{\rm B}$; (b) distance criterion with 
$r_{\rm c}=\lambda_{\rm B}/2$; (c) distance criterion with $r_{\rm 
c}=2\sigma$; (d) fitting Eq.~(\ref{eqn:prd}) to $p(r)$.}

\end{figure}

Figure \ref{fig:alpha} shows the degree of association $\alpha$ as a 
function of reduced ion density $\rho^{*}$ along several isotherms, 
$T^{*}=0.04$, $0.05$, $0.06$, and $0.07$. Recall that the critical 
temperature of the RPM is $T_{\rm c}^{*} \simeq 0.05$. Four sets of 
simulation data are shown, corresponding to different clustering 
criteria. The simulation data in Figs.~\ref{fig:alpha}(a)-(c) were 
obtained using distance-based criteria of $r_{\rm c}=\lambda_{\rm B}$, 
$\lambda_{\rm B}/2$, and $2\sigma$; the data in Fig.~\ref{fig:alpha}(d) 
were obtained from fits to $p(r)$. Also included in the figures are the 
theoretical predictions of Eq.~(\ref{eqn:alpha}) with $K$ evaluated 
numerically using Eq.~(\ref{eqn:k}) and $r_{c}=\lambda_{\rm B}/2$, and 
with the asymptotic expression in Eq.~(\ref{eqn:kasymp}). On the scale 
of these plots, curves with $r_{\rm c}=\lambda_{\rm B}$ are 
indistinguishable from those with $r_{\rm c}=\lambda_{\rm B}/2$ and so 
they are omitted.

%%% Figure 1 %%%

The first impression given by Fig.~\ref{fig:alpha} is that there is very 
good overall agreement between the simulation results and the 
theoretical predictions. A close inspection of 
Figs.~\ref{fig:alpha}(a)-(c) shows that, in simulations, the distance 
criteria $r_{\rm c}=\lambda_{\rm B}$ and $\lambda_{\rm B}/2$ give 
slightly poorer results for the degree of association; looking at 
Figs.~\ref{fig:alpha}(a) and (b), the data along the higher temperature 
isotherms vary a little too sharply and saturate at $\alpha=1$ 
prematurely as the density is increased. Figure \ref{fig:alpha}(c) shows 
that the fixed-distance cut-off of $r_{\rm c}=2\sigma$ provides a more 
realistic variation with density, reflecting a strong association of 
ions in clusters close to contact.

\begin{figure}[tb]

\includegraphics[width=8.5cm]{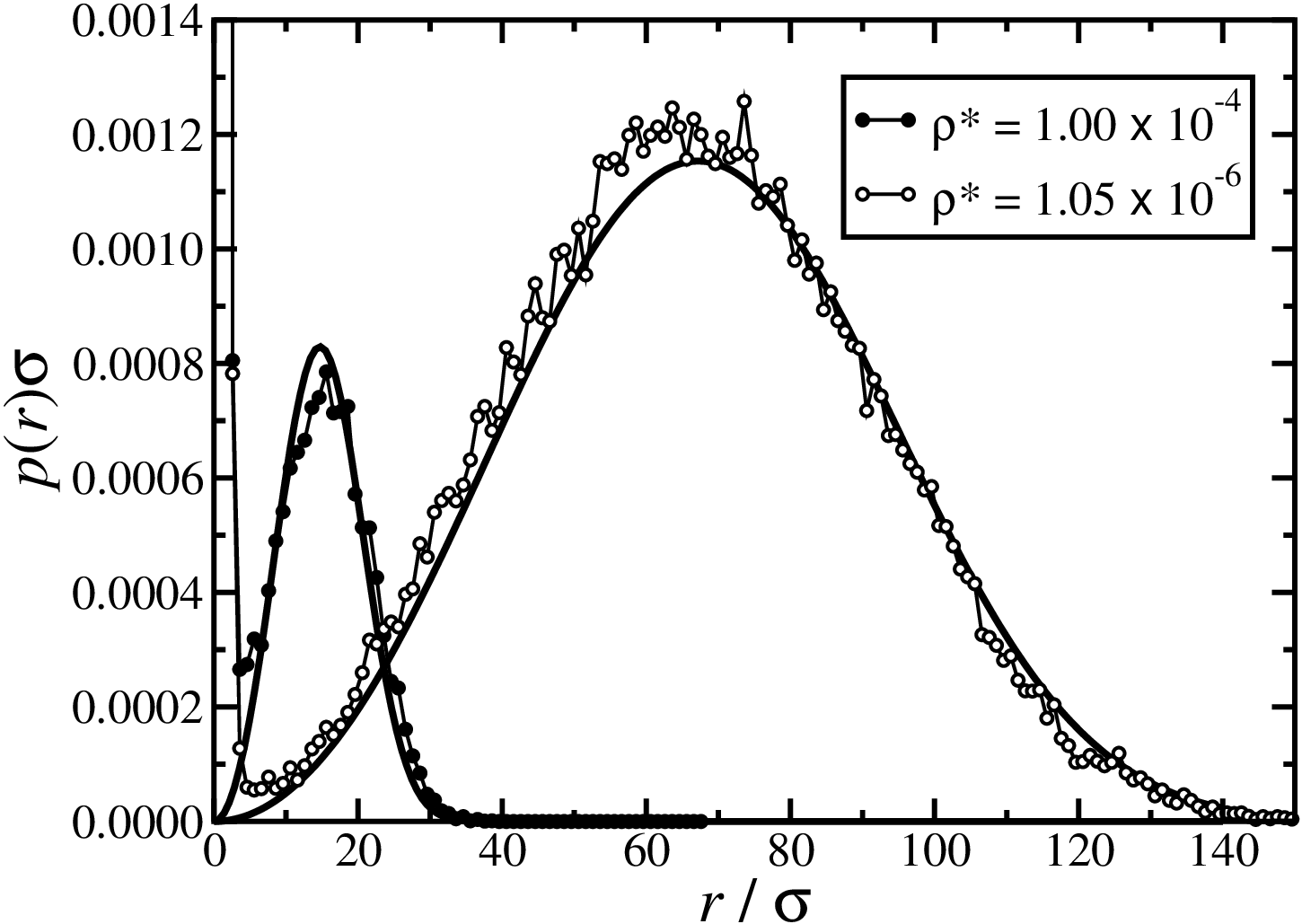}

\caption{\label{fig:pr} Nearest-neighbour distribution function $p(r)$ 
at $T^{*}=0.05$. The simulation results are from simulations with 
$\rho^{*}=1.05 \times 10^{-6}$ (open symbols) and $\rho^{*}=1.00 \times 
10^{-4}$ (filled symbols); the thick curves are fits using the 
dissociated-ion result in Eq.~(\ref{eqn:prd}) over the range $r>r_{0}$, 
where $r_{0}$ is the maximum in $p(r)$.}

\end{figure}

Further insights are afforded by simulation measurements of the 
nearest-neighbour cation-anion distribution function, $p(r)$. Two 
examples from the $T^{*}=0.05$ isotherm are shown in Fig.~\ref{fig:pr}, 
at densities of $\rho^{*}=1.05 \times 10^{-6}$ and $1.00 \times 
10^{-4}$. The key point is that at these low densities, $p(r)$ appears 
to be a superposition of two parts: a short-range associated-ion 
contribution, which dies off by about $r=2$-$3\sigma$; and a peaked 
contribution corresponding to free ions. The short-range part decays 
within a distance much shorter than the Bjerrum length commonly used as 
a distance-based clustering criterion, which at this temperature is 
$20\sigma$. Assuming no correlations between free ions and any other 
ions in the system, the peaks in $p(r)$ should occur at $r_{0} \simeq 
67\sigma$ and $15\sigma$ for $\rho^{*}=1.05 \times 10^{-6}$ and $1.00 
\times 10^{-4}$, respectively; by comparison with the simulation 
results, these predictions are very reliable. The free-ion peaks are 
very broad, showing that a distance-based criterion $r_{\rm c} \sim 
\lambda_{\rm B}$ is not physically justified. \cite{Allahyarov:2007/a} 
As the density of ions is increased, the peak both shifts to lower 
values of $r$ and decreases in height. Equation (\ref{eqn:prd}) provides 
excellent fits to the simulation results for $p(r)$ (for $r > r_{0}$), 
and yields values for the degree of association $\alpha$ as shown in 
Fig.~\ref{fig:alpha}(d). There is very good agreement with the 
simulation results using $r_{\rm c}=2\sigma$, in correspondence with the 
comments made above regarding the decomposition of $p(r)$ in to a 
short-ranged associated-ion contribution and a broad free-ion 
contribution. Accordingly, there is excellent agreement between the 
theoretical predictions for $\alpha$ and the results of the analysis of 
$p(r)$.

%%% Figure 2 %%%

\begin{figure}[b]

\includegraphics[width=8.5cm]{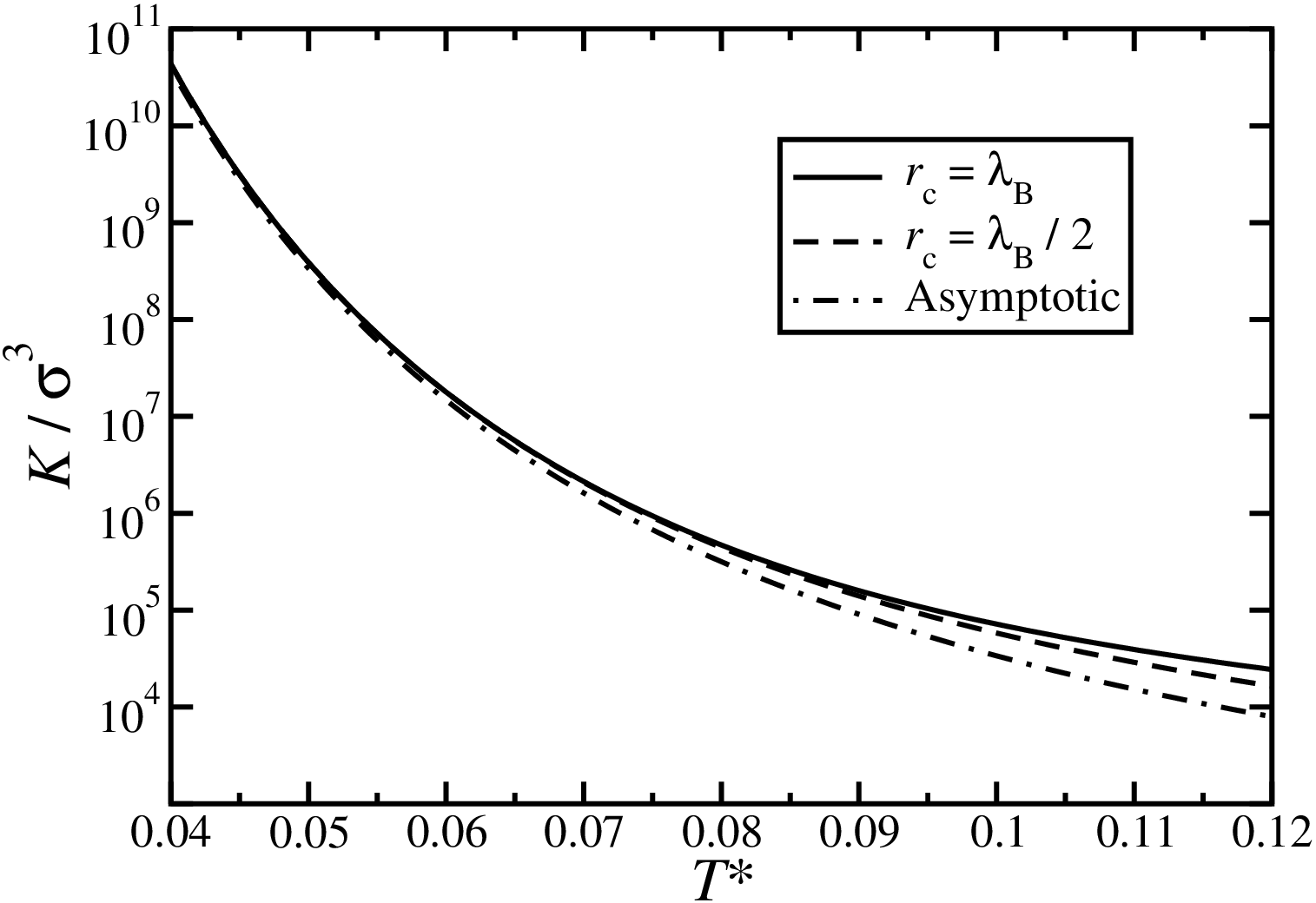}

\caption{\label{fig:k} Cation-anion configurational integral, $K$, as a 
function of temperature $T^{*}$: Eq.~(\ref{eqn:k}) with $r_{\rm 
c}=\lambda_{\rm B}$ (solid line); Eq.~(\ref{eqn:k}) with $r_{\rm 
c}=\lambda_{\rm B}/2$ (dashed line); Eq.(\ref{eqn:kasymp}) (dot-dashed 
line).}

\end{figure}

Figure \ref{fig:alpha} shows that the theoretical predictions are not 
very sensitive to the precise values of $r_{\rm c}$ and hence $K$. This 
is explored further in Fig.~\ref{fig:k}, which shows $K$ as a function 
of temperature evaluated using Eq.~(\ref{eqn:k}) with $r_{\rm 
c}=\lambda_{\rm B}$ and $\lambda_{\rm B}/2$, and from the asymptotic 
expression in Eq.~(\ref{eqn:kasymp}). The first two expressions give 
essentially identical numerical results over the temperature range $0.04 
\leq T^{*} \leq 0.10$, the region of current interest. The asymptotic 
expression, Eq.~(\ref{eqn:kasymp}), is accurate only at the lower end of 
the temperature range. Note that $K$ is plotted on a logarithmic scale; 
the deviation between the `Bjerrum length' and asymptotic results at 
$T^{*}=0.07$ is about 20\%.

%%% Figure 3 %%%

\begin{figure}[tb]

\includegraphics[width=8.5cm]{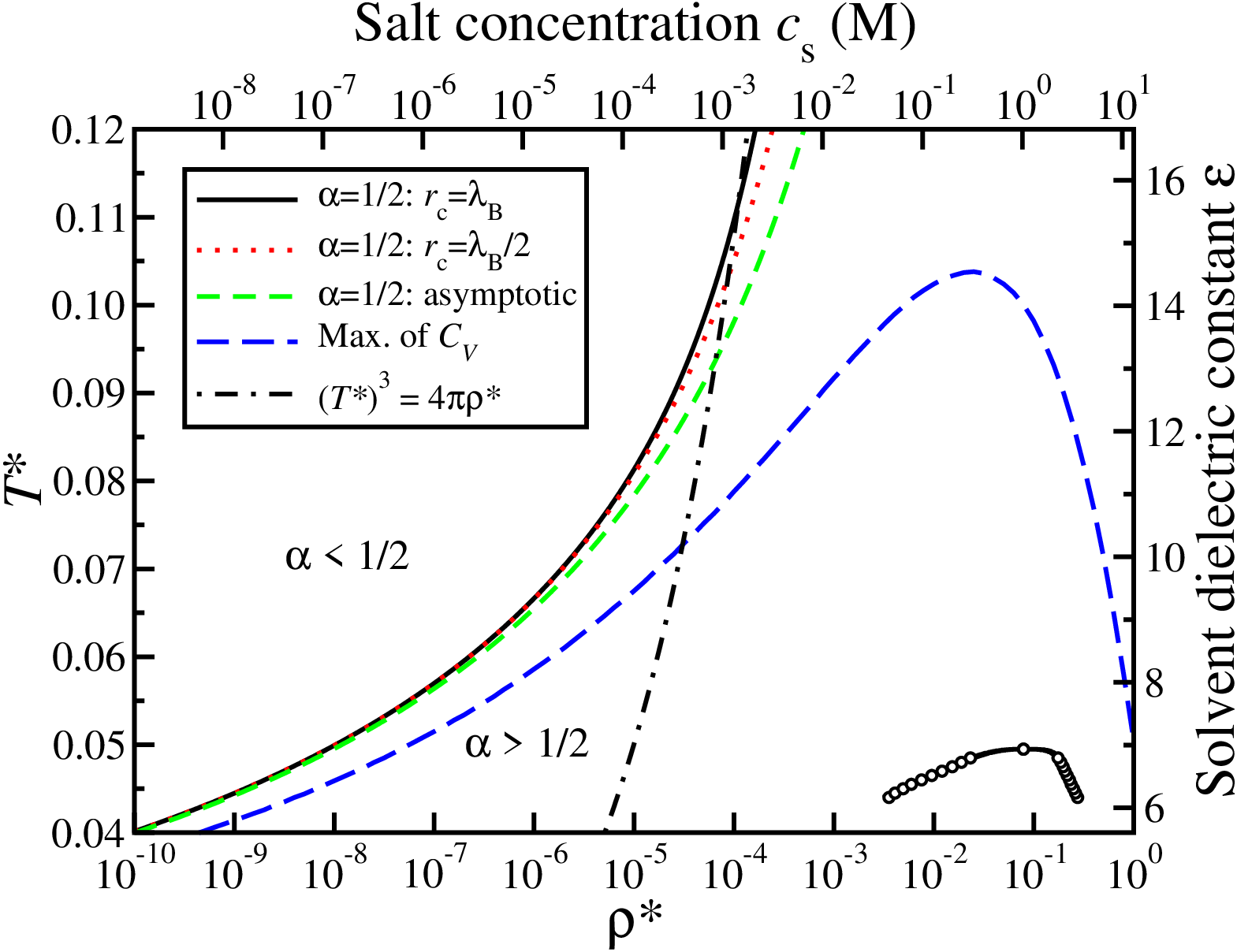}

\caption{\label{fig:locus} Phase diagram of the RPM showing the 
vapour-liquid coexistence region (lower right) from simulations, 
\cite{Romero:2002/a} and the boundary between dissociated ($\alpha < 
\frac{1}{2}$) and associated ($\alpha > \frac{1}{2}$) regimes as 
predicted from Eq.~(\ref{eqn:bjlocus}) with three different evaluations 
of $K$: Eq.~(\ref{eqn:k}) with $r_{\rm c}=\lambda_{\rm B}$ (black solid 
line) and $r_{\rm c}=\lambda_{\rm B}/2$ (red dotted line); 
Eq.~(\ref{eqn:kasymp}) (green short-dashed line). Also shown is the 
locus of maxima in the constant-volume heat capacity $C_{V}$ from a 
simple two-particle theory \cite{PJC:1999/a} (blue long-dashed line) and 
the boundary between ideal and strongly correlated regimes predicted by 
DH theory from Eq.~(\ref{eqn:dhlocus}) (black dot-dashed line). Real 
units are shown for a monovalent salt with ionic diameter 
$\sigma=4~\mbox{\AA}$ at a temperature $T=298.15~\mbox{K}$.}

\end{figure}

When considering the effective interactions between charged colloids, it 
is of primary importance to know the degree of association of 
counterions and added salt within the suspending phase. 
\cite{Leunissen:2007/b,Zwanikken:2007/a,Allahyarov:2007/a,% 
Allahyarov:2008/a,Zwanikken:2009/a} Of course, detailed calculations are 
easily performed using the prescriptions outlined herein. For a 
qualitative assessment, however, it is useful to divide the phase 
diagram into regions where the ions are mostly dissociated ($\alpha < 
\frac{1}{2}$) and where they are mostly associated ($\alpha > 
\frac{1}{2}$). Within the theory outlined in Section \ref{sec:theory}, 
the dividing line is defined by Eq.~(\ref{eqn:bjlocus}). The phase 
diagram in the $\rho^{*}$-$T^{*}$ plane is shown in 
Fig.~\ref{fig:locus}; the vapour-liquid coexistence data are taken from 
Ref.~\onlinecite{Romero:2002/a}. The $\alpha=\frac{1}{2}$ line is shown 
for the three expressions for $K$, with $r_{\rm c}=\lambda_{\rm B}$ and 
$\lambda_{\rm B}/2$, and from Eq.~(\ref{eqn:kasymp}). The deviations 
between these expressions become more pronounced as temperature 
increases, due to the increasingly significant large-$r$ contributions 
to the integral in Eq.~(\ref{eqn:k}).

%%% Figure 4 %%%

It should be noted that significant improvements on the non-interacting 
particle theory are possible, and indeed have been developed in detail. 
Already in 1926, Bjerrum took account of the reduction in free-ion 
concentration in order to compute the mean activity coefficients in 
electrolyte solutions. \cite{Bjerrum:1926/a,Fowler:1956/a} Fisher and 
Levin have explored the consequences of this level of approximation on 
the thermodynamics and phase behaviour of the RPM in their 
Debye-H{\"u}ckel-Bjerrum theory, and found that it leads to a 
vapour-liquid coexistence curve of the incorrect shape. 
\cite{Fisher:1993/a,Levin:1996/a} Including ion-ion pair (ion-dipole) 
interactions restores the correct shape of the coexistence curve, and 
yields quite accurate values for the critical parameters. These 
extensions are not applied here to the problem of ion pairing: as 
Figs.~\ref{fig:alpha}(c) and (d) show, the agreement between theory and 
simulation at the low concentrations and low temperatures of interest is 
excellent; the simplest ion-pairing theory is clearly adequate for the 
present purposes.

The DH expression for the osmotic pressure of the electrolyte, in RPM 
reduced units, reads \cite{Debye:1923/a,McQuarrie:1976/a,Levin:1996/a}
\begin{eqnarray}
\frac{\Pi\sigma^{3}}{k_{\rm B}T} 
&=&       \frac{1}{4\pi}
          \left[ x^{2}T^{*} + \ln{(1+x)} - x + \frac{x^{2}}{2(1+x)} \right]
          \nonumber \\
&\approx& \rho^{*}\left( 1 - \frac{x}{6T^{*}} + \ldots \right)
\end{eqnarray}
where $x=\kappa_{\rm D}\sigma=\sqrt{4\pi\rho^{*}/T^{*}}$ and 
$\kappa_{\rm D}^{-1}$ is the Debye screening length. This expression is 
exact to leading order in $x$ as $\rho^{*}\rightarrow 0$. At low density 
and high temperature, the ions are largely dissociated and the 
thermodynamics is essentially ideal. Significant deviations from 
ideality are expected when $x/T^{*} \sim 1$, and a dividing line between 
free-ion and strongly correlated regimes can therefore be defined by
\begin{equation}
T^{*} = \sqrt[3]{4\pi\rho^{*}}.
\label{eqn:dhlocus}
\end{equation}
This line is included in Fig.~\ref{fig:locus}, and shows that there is a 
significant portion of the phase diagram in which the DH theory would 
suggest a low degree of ion association (because $x/T^{*} < 1$), but the 
simulations and the Bjerrum theory show that $\alpha > \frac{1}{2}$. 
This has a serious consequence for linearised Poisson-Boltzmann theories 
of electrolyte solutions and related systems, which assume weak ion-ion 
correlations; in that part of the phase diagram lying between the lines 
defined by Eqs.~(\ref{eqn:bjlocus}) and (\ref{eqn:dhlocus}), ions are 
associated and hence strongly correlated despite the fact that $x/T^{*} 
< 1$.

There is one more feature of the $\alpha=\frac{1}{2}$ locus to be 
discussed, and that is its monotonic variation with density. At high 
enough density, the distinction between two free ions and one ion pair 
becomes blurred. At the simplest level, this volume effect can be 
captured by a two-particle theory, in which all ions are resolved into 
cation-anion pairs, and each ion pair has an internal configurational 
integral given by
\begin{equation}
q_{2} = 4\pi\int_{\sigma/2}^{s_{\rm c}}
        s^{2}\exp{\left(\frac{\sigma}{2sT^{*}}\right)}
        {\rm d}s
\end{equation}
where $s$ is the distance from the ion-pair center of mass to one of the 
constituent ions, and the upper limit $s_{\rm c}=(3/2\pi\rho)^{1/3}$ 
fixes the volume per pair to be $2/\rho$. Although the degree of 
association is not defined within this theory, one can delineate the 
boundary between dissociated and associated regimes with the locus of 
maxima in the constant-volume heat capacity $C_{V} = k_{\rm 
B}\beta^{2}(\partial^{2}\ln{q_{2}}/\partial\beta^{2})_{V}$, where 
$\beta=1/k_{\rm B}T$. \cite{PJC:1999/a} This line is shown in 
Fig.~\ref{fig:locus} and suggests that the domain of associated ions is 
bounded from above ($T^{*} \simeq 0.1$). This is in correspondence with 
the types of phase diagrams proposed for a wide range of ionic fluids. 
\cite{Pitzer:1990/a} It was shown in earlier work that the two-particle 
theory provides an excellent account of simulation measurements for the 
maxima in $C_{V}$. \cite{PJC:1999/a}

\subsection{Analysis of experimental data} \label{sec:exp_results}

To aid comparisons with experimental systems, axes in 
Fig.~\ref{fig:locus} are also shown with real units for an electrolyte 
with ionic diameter $\sigma=4~\mbox{\AA}$ at a temperature 
$T=298.15~\mbox{K}$. The concentration $c_{\rm s}$ is given in moles of 
salt (not ions) per litre, and with the physical temperature held 
constant, $T^{*}$ becomes proportional to the solvent dielectric 
constant $\epsilon$.

To effect a direct comparison between experiment and theory (and 
therefore to link experiment with simulation), attention is turned to 
the degree of association, $\alpha$. This can be extracted from 
experimental measurements of the molar conductivity. Ignoring the 
formation of ion triples and higher charged clusters, $\alpha = 
1-\Lambda/\Lambda_{0}$ where $\Lambda$ is the molar conductivity and 
$\Lambda_{0}$ is its limiting value at infinite dilution. The most 
common source of experimental uncertainty is in the determination of 
$\Lambda_{0}$, since a suitable model has to be used to extrapolate 
$\Lambda$ to infinite dilution. For the present purposes, conductivity 
data for tetraalkylammonium salts in various solvents are analysed using 
the quoted estimates for $\Lambda_{0}$. In particular, Leunissen and 
co-workers have obtained the degree of association of tetrabutylammonium 
halides in bromocyclohexane and decalin-bromocyclohexane mixtures. 
\cite{Leunissen:2007/c,Hollingsworth:2010/a} Conductivity data for 
tetrapropylammonium picrate in chlorobenzene \cite{Lindback:1980/a} and 
tetrabutylammonium iodide in carbon tetrachloride-nitrobenzene mixtures 
\cite{Roy:2009/a} are also analysed. The system parameters are 
summarised in Table \ref{tab:experiment}; the experimental data span 
wide ranges of concentration and solvent dielectric constant.

% Table %
\begin{table*}[t]

\caption{\label{tab:experiment} Physical parameters for 
tetraalkylammonium salts in various solvents: $\epsilon$ is the solvent 
dielectric constant, $\lambda_{\rm B}$ is the Bjerrum length for 
$T=298.15~\mbox{K}$, $c_{\rm s}$ is the salt concentration, $K_{\rm M}$ 
is the dimensionless ion-pairing association constant (defined on the 
molar scale), and $\sigma$ is the effective hard-sphere diameter of the 
ion. $T^{*}=\sigma/\lambda_{\rm B}$ and $\rho^{*}$ are the effective RPM 
temperature and density, respectively. ($\mbox{Bu}=$butyl, 
$\mbox{Pr}=$propyl, $\mbox{Pi}=$picrate.)}

\begin{tabular}{lcccccccc}\hline\hline

System & 
Ref.\ & 
$\epsilon$ & 
$\lambda_{\rm B}$ (\AA) & 
$c_{\rm s}$ ($\mu\mbox{M}$) & 
$K_{\rm M}$ & 
$\sigma$ (\AA) & 
$T^{*}$ & 
$10^{6}\rho^{*}$ \\ \hline

$\mbox{Bu}_{4}\mbox{N}^{+}\mbox{Cl}^{-}$ / 
$\mbox{C}_{6}\mbox{H}_{11}\mbox{Br}$ & \onlinecite{Leunissen:2007/c} &
$7.92$ &
$70.8$ & 
$0.28$--$244$ & 
$3.80 \times 10^{7}$ &
$3.27$ & 
$0.0462$ &
$0.0118$--$10.3$ \\

$\mbox{Bu}_{4}\mbox{N}^{+}\mbox{Br}^{-}$ / 
$\mbox{C}_{6}\mbox{H}_{11}\mbox{Br}$ & \onlinecite{Hollingsworth:2010/a} 
&
$7.92$ &
$70.8$ & 
$0.56$--$245$ &
$9.17 \times 10^{6}$ & 
$3.57$ &
$0.0504$ &
$0.0307$--$13.4$ \\

$\mbox{Bu}_{4}\mbox{N}^{+}\mbox{Br}^{-}$ / 27.3 wt\% 
decalin-$\mbox{C}_{6}\mbox{H}_{11}\mbox{Br}$ & 
\onlinecite{Hollingsworth:2010/a} &
$5.62$ & 
$99.7$ & 
$0.47$--$157$ & 
$6.31 \times 10^{6}$ & 
$5.53$ & 
$0.0555$ & 
$0.0957$--$32.0$ \\

$\mbox{Pr}_{4}\mbox{N}^{+}\mbox{Pi}^{-}$ / $\mbox{PhCl}$ & 
\onlinecite{Lindback:1980/a} &
$5.612$ & 
$99.9$ & 
$5.97$--$1539$ & 
$8.80 \times 10^{6}$ & 
$5.41$ & 
$0.0542$ & 
$1.14$--$294$ \\

$\mbox{Bu}_{4}\mbox{N}^{+}\mbox{I}^{-}$ / 80 wt\% 
$\mbox{CCl}_{4}$-$\mbox{PhNO}_{2}$ & \onlinecite{Roy:2009/a} &
$10.22$ & 
$54.8$ & 
$170$--$1330$ & 
$6.44 \times 10^{5}$ & 
$3.16$ & 
$0.0576$ & 
$6.46$--$50.5$ \\

$\mbox{Bu}_{4}\mbox{N}^{+}\mbox{I}^{-}$ / 60 wt\% 
$\mbox{CCl}_{4}$-$\mbox{PhNO}_{2}$ & \onlinecite{Roy:2009/a} &
$17.45$ & 
$32.1$ & 
$170$--$1320$ & 
$1.23 \times 10^{3}$ &
$3.17$ & 
$0.0987$ & 
$6.52$--$50.6$ \\

$\mbox{Bu}_{4}\mbox{N}^{+}\mbox{I}^{-}$ / 40 wt\% 
$\mbox{CCl}_{4}$-$\mbox{PhNO}_{2}$ & \onlinecite{Roy:2009/a} &
$23.90$ &
$23.5$ &
$440$--$4070$ &
$2.91 \times 10^{2}$ &
$2.60$ &
$0.111$ &
$9.31$--$86.2$ \\

$\mbox{Bu}_{4}\mbox{N}^{+}\mbox{I}^{-}$ / 20 wt\% 
$\mbox{CCl}_{4}$-$\mbox{PhNO}_{2}$ & \onlinecite{Roy:2009/a} &
$29.66$ &
$18.9$ &
$830$--$7380$ &
$1.57 \times 10^{2}$ & 
$2.08$ &
$0.110$ &
$9.00$--$80.0$ \\ \hline\hline

\end{tabular}
\end{table*}

\begin{figure}[tb]

\includegraphics[width=8.5cm]{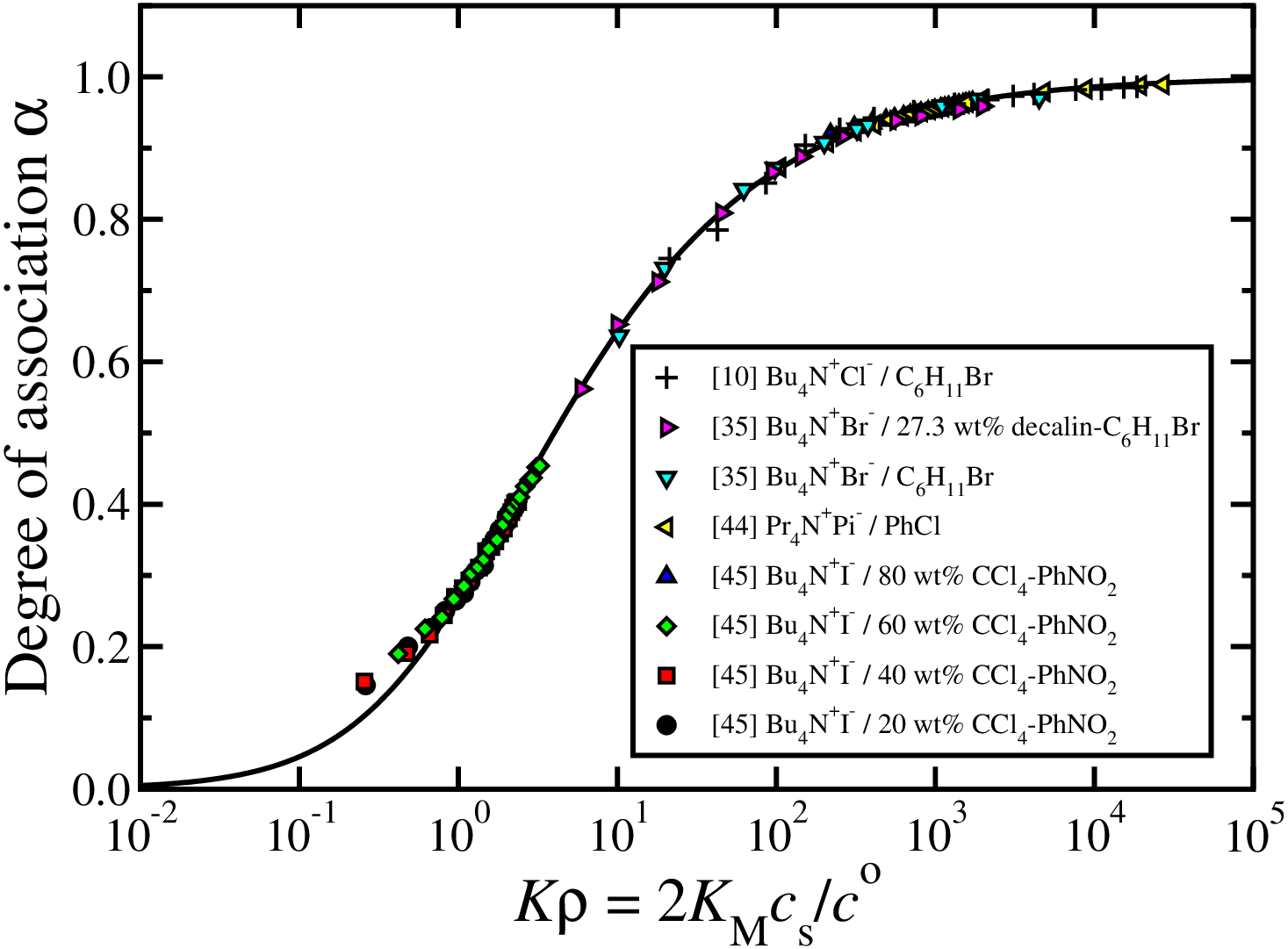}

\caption{\label{fig:experiment} Degree of association $\alpha$ against 
$K\rho = 2K_{\rm M}c_{\rm s}/c^{\minuso}$ for some tetraalkylammonium 
salts in various solvents. References for the experimental data are 
given in square brackets. The solid line is Eq.~(\ref{eqn:alpha}). 
($\mbox{Bu}=$butyl, $\mbox{Pr}=$propyl, $\mbox{Pi}=$picrate.)}

\end{figure}

The experimental results for $\alpha$ are fitted using 
Eq.~(\ref{eqn:alpha}), with the association constant as a fitting 
parameter. In fact, for the purposes of analysis, $K\rho$ is replaced by 
$2K_{\rm M}c_{\rm s}/c^\minuso$, where $K_{\rm M}$ is the dimensionless 
ion-pairing equilibrium constant (defined on the molar scale), $2c_{\rm 
s}$ is the ion concentration in $\mbox{M}$, and $c^\minuso=1~\mbox{M}$. 
The fitted values of $K_{\rm M}$ are reported in Table 
\ref{tab:experiment}. The experimental values for $\alpha$ are shown in 
Fig.~\ref{fig:experiment}, plotted against $K\rho = 2K_{\rm M}c_{\rm 
s}/c^\minuso$. There is an impressive collapse of the experimental data 
on to the theoretical universal curve given by Eq.~(\ref{eqn:alpha}).

%%% Figure 5 %%%

It is possible to convert the experimental parameters in to RPM units. 
Since $q$, $\epsilon$, $T$, and hence $\lambda_{\rm B}$ are all known, 
an effective ion diameter $\sigma$ can be obtained by equating the 
fitted values of $K_{\rm M}$ with the expression for $K$ in 
Eq.~(\ref{eqn:k}), the numerical conversion between the two being
\begin{equation}
K = \frac{K_{\rm M}}{1000N_{\rm A}c^\minuso}
\end{equation}
where $N_{\rm A}$ is Avogadro's number. The integrand in 
Eq.~(\ref{eqn:k}) is $r^{2}\exp{(\lambda_{\rm B}/r)}$, and $r_{\rm 
c}=\lambda_{\rm B}/2$. (Other sensible choices for $r_{\rm c}$ give 
essentially the same numerical results.) The ion diameter $\sigma$ is 
obtained by numerical solution of Eq.~(\ref{eqn:k}). This procedure 
ignores chemical detail such as the precise nature of the short-range 
repulsive interactions and the presence of van der Waals interactions, 
and so the fitted values of $\sigma$ might reflect some non-coulombic 
effects. Nonetheless, under the conditions considered here, the 
coulombic interaction between ions at contact is dominant (being in the 
range of $10$-$20~k_{\rm B}T$) and so the effective RPM parameters are 
expected to be meaningful.

Values of $\lambda_{\rm B}$, $\sigma$, $T^{*}=\sigma/\lambda_{\rm B}$, 
and the range of RPM ion densities $\rho^{*}$ are all summarised in 
Table \ref{tab:experiment}. The data of Leunissen and co-workers 
correspond to RPM temperatures in the region of $T^{*}=0.05$, and RPM 
ion densities as low as $\rho^{*}=10^{-8}$; Lindb{\"a}ck and Beronius' 
data \cite{Lindback:1980/a} correspond to a similar temperature, but do 
not extend to as low concentration. The data of Roy {\it et al.} 
\cite{Roy:2009/a} do not extend to very low concentration, but they do 
span the temperature range $0.06 \leq T^{*} \leq 0.11$. Note that the 
experimental data correspond to the regime where the mean ion-ion 
separation ($\rho^{-1/3}$) is much greater than the Bjerrum length, 
i.e., in RPM units $\sqrt[3]{4\pi\rho^{*}} \ll T^{*}$. This corresponds 
to the weakly interacting regime of the DH theory summarised in Section 
\ref{sec:rpm_results}, and so the theory would not have predicted the 
extensive pairing apparent in experiments. The central conclusion is 
that the Bjerrum theory (and by association, the simulations) 
successfully treats the ion-pairing equilibrium of nanomolar salt 
solutions in low-polarity solvents, such as those employed in recent 
work on charged-colloid suspensions. 
\cite{Yethiraj:2003/a,Royall:2003/a,Leunissen:2005/a,% 
Campbell:2005/a,Shevchenko:2006/a,Lu:2006/a,Leunissen:2007/a,% 
Leunissen:2007/b,Leunissen:2007/c,Zwanikken:2007/a}

The effects of ion pairing on the effective screening length will now be 
considered. The appropriate effective screening parameter in the 
presence of association is clearly
\begin{equation}
\kappa = \sqrt{4\pi(1-\alpha)\rho\lambda_{\rm B}}
       = \kappa_{\rm D}\sqrt{1-\alpha}
\label{eqn:kappaeff}
\end{equation}
where $\kappa_{\rm D}=\sqrt{4\pi\rho\lambda_{\rm B}}$ is the inverse of 
the Debye length. At very low salt concentrations where $K\rho\ll 1$, 
Eq.~(\ref{eqn:alpha}) shows that $\alpha \approx 
\mbox{$\frac{1}{2}$}K\rho$, and hence
\begin{equation}
\frac{\kappa}{\kappa_{\rm D}} \approx 1-\mbox{$\frac{1}{4}$}K\rho.
\label{eqn:kappa_lo}
\end{equation} 
The screening length is then essentially equal to that assuming complete 
dissociation, i.e., the Debye length. At higher concentrations where 
$K\rho\gg 1$, Eq.~(\ref{eqn:alpha}) predicts $\alpha \approx 
1-\sqrt{2/K\rho}$ and hence
\begin{equation}
\frac{\kappa}{\kappa_{\rm D}} \approx 
\left(\frac{2}{K\rho}\right)^{1/4}.
\label{eqn:kappa_hi}
\end{equation}
In this regime, $\kappa \ll \kappa_{\rm D}$ and hence the screening 
length is much longer than would be expected on the basis of complete 
dissociation. Moreover, there is an unusual scaling behaviour: because 
$\kappa_{\rm D}$ has a $\rho^{1/2}$ dependence, the effective scaling 
parameter scales like $\kappa \sim \rho^{1/2}\cdot\rho^{-1/4} = 
\rho^{1/4}$. Hence, the effective screening length scales like 
$\kappa^{-1} \sim \rho^{-1/4}$, while the Debye length scales like 
$\kappa_{\rm D}^{-1} \sim \rho^{-1/2}$.

\begin{figure}[b]

\includegraphics[width=8.5cm]{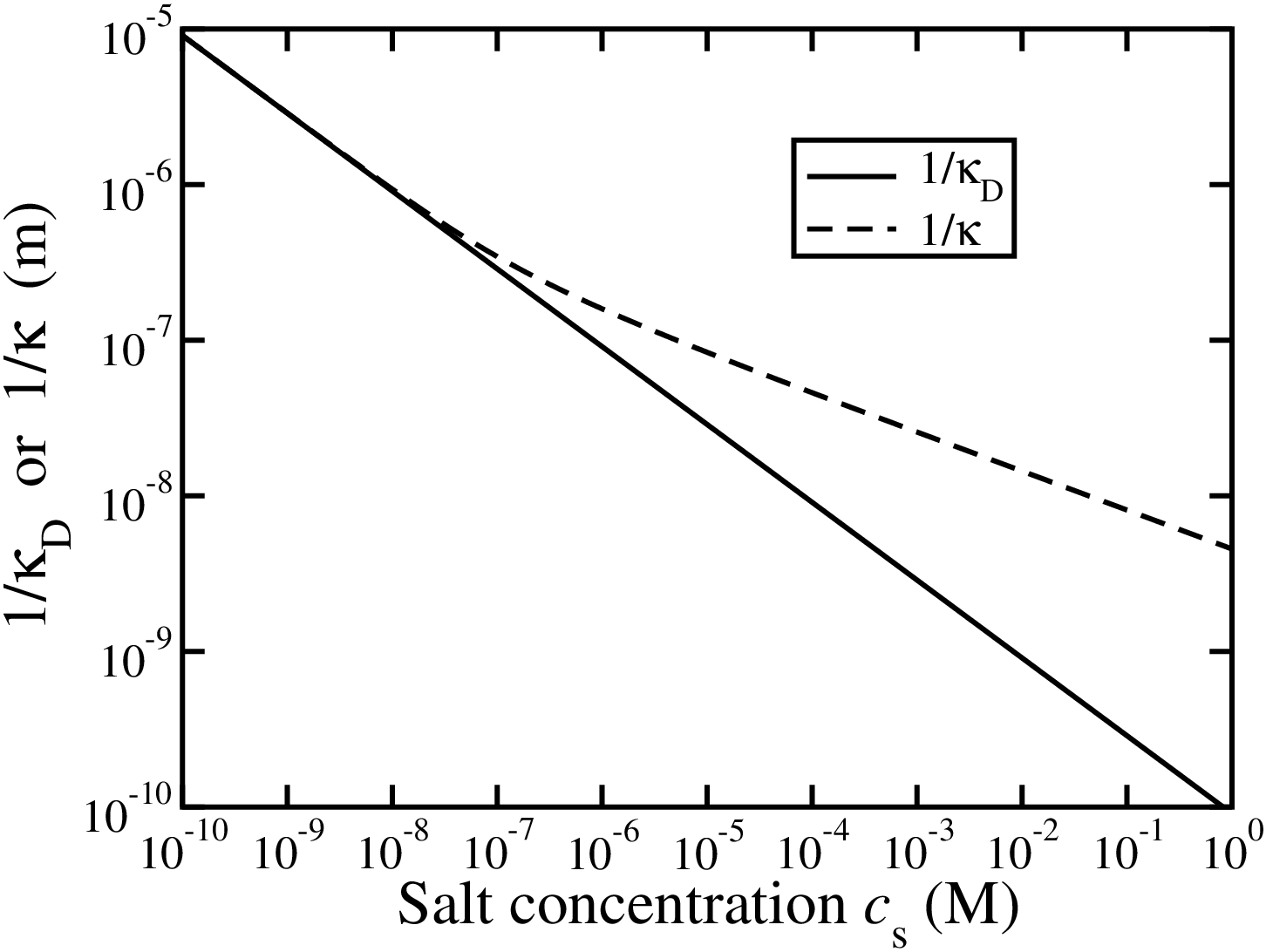}

\caption{\label{fig:lengths} The Debye length $\kappa_{\rm D}^{-1}$, and 
the effective screening length $\kappa^{-1}$ from 
Eq.~(\ref{eqn:kappaeff}), for a solution with $\sigma=4~\mbox{\AA}$, 
$\epsilon=7$, and $T=298.15~\mbox{K}$; this set of parameters 
corresponds to a reduced RPM temperature of $T^{*}=0.05$.}

\end{figure}

A glance at Fig.~\ref{fig:experiment} shows that the regime $K\rho \gg 
1$ is in fact experimentally accessible. In particular, the experimental 
measurements by Leunissen and co-workers 
\cite{Leunissen:2007/c,Hollingsworth:2010/a} extend to a very high 
degree of association, and so the resulting interactions between charged 
colloids suspended in these solutions will not be screened as 
effectively as might be expected. As an example, Fig.~\ref{fig:lengths} 
shows the Debye length $\kappa_{\rm D}^{-1} \propto c_{\rm s}^{-1/2}$, 
and the effective screening length $\kappa^{-1}$ from 
Eq.~(\ref{eqn:kappaeff}), for a solution with $\sigma=4~\mbox{\AA}$, 
$\epsilon=7$, and $T=298.15~\mbox{K}$; this set of parameters 
corresponds to a reduced RPM temperature of $T^{*}=0.05$, and is 
representative of experiments in low-polarity solvents. Over the range 
$10^{-9}~\mbox{M} \leq c_{\rm s} \leq 10^{-7}~\mbox{M}$, the screening 
length and the Debye length coincide, ranging from several micrometres 
down to several hundred nanometres; these values correspond well with 
experimentally determined values. 
\cite{Leunissen:2007/a,Leunissen:2007/b} At higher salt concentrations 
$c_{\rm s} > 10^{-7}~\mbox{M}$, $\kappa^{-1}$ exceeds $\kappa_{\rm 
D}^{-1}$ due to the formation of ion pairs. In addition, $\kappa^{-1}$ 
decays less fast with increasing concentration, switching over to the 
$c_{\rm s}^{-1/4}$ dependence advertised in Eq.~(\ref{eqn:kappa_hi}). 
Finally, it is noted that at very high concentrations (outwith the 
relevant range studied here) the formation of ion pairs can lead to 
$\kappa^{-1}$ increasing with increasing $c_{\rm s}$; the ion-pair 
contribution to the effective dielectric constant of the solution 
increases with increasing concentration, which ultimately leads to 
reductions in $\lambda_{\rm B}$ and $\kappa$. \cite{Zwanikken:2009/a}

%%% Figure 6 %%%

Some interesting transient behaviour has been observed in experiments on 
charged-colloid suspensions. \cite{Royall:2003/a} Specifically, the 
colloids show the effects of anomalously long-ranged repulsions which 
are incompatible with the apparent salt concentration $c_{\rm s} \sim 
10^{-8}~\mbox{M}$. This transient behaviour can occur on the timescale 
of a few days. One contributing factor might be the time taken for ions 
to associate and/or dissociate after preparation. For instance, if ions 
were initially associated, then the screening length and colloidal 
repulsions would decrease on the approach to equilibrium. Assuming that 
the association of cations with anions is a second-order, 
diffusion-controlled `reaction', then the corresponding macroscopic rate 
constant can be estimated by $k_{\rm a}=8RT/3\eta$, \cite{Atkins:2010/a} 
where $\eta$ is the viscosity of the solvent. For a solvent with $\eta = 
10^{-3}~\mbox{Pa}~\mbox{s}$ at room temperature, this yields $k_{\rm a} 
= 7 \times 10^{9}~\mbox{M}^{-1}~\mbox{s}^{-1}$. The corresponding 
first-order dissociation of ion pairs will have a rate constant $k_{\rm 
d}=k_{\rm a}{c^\minuso}/K_{\rm M}$. Table \ref{tab:experiment} shows 
that $K_{\rm M}$ can be as high as $10^{7}$, and so the characteristic 
time for ion dissociation will be no more than $1/k_{\rm d} \sim 
0.001~\mbox{s}$. This is only a very rough estimate and solvation-shell 
structure and the slow escape from the long-range coulombic attraction 
between ions may well increase this timescale considerably, but it seems 
unlikely that ion dissociation is the dominant cause of the transient 
behaviour. A more prosaic explanation for the observed transients in 
Ref.~\onlinecite{Royall:2003/a} may be that ions are initially 
sequestered but subsequently released by impurities such as water or by 
the container walls, leading to a slow reduction in the screening 
length.

\section{Conclusions} \label{sec:conclusions}

In this paper, the association of ions in low-polarity solvents was 
studied within the context of the restricted primitive model at low 
concentration and low temperature. In reduced units, concentrations as 
low as $10^{-10}$ and temperatures as low as $0.04$ were simulated using 
an efficient Monte Carlo algorithm, \cite{PJC:2010/b} and comparisons 
were made with Bjerrum's quasi-chemical ion-pairing theory. 
\cite{Bjerrum:1926/a} These conditions correspond to nanomolar salt 
solutions in low dielectric constant solvents under ambient conditions; 
such media are of relevance to recently synthesised charged-colloid 
suspensions. \cite{Yethiraj:2003/a,Royall:2003/a,Leunissen:2005/a,% 
Campbell:2005/a,Shevchenko:2006/a,Lu:2006/a,Leunissen:2007/a,% 
Leunissen:2007/b,Leunissen:2007/c,Zwanikken:2007/a} The degree of ion 
association strongly affects the effective interactions between 
colloids, \cite{Allahyarov:2007/a,Allahyarov:2008/a,Zwanikken:2009/a} 
and so one of the aims of this work was to map out the different regimes 
of association on the phase diagram.

In the simulations, distance-based criteria and the nearest-neighbour 
cation-anion distribution function $p(r)$ were used to determine the 
degree of association. It was shown that conventional distance-based 
criteria based on the Bjerrum length \cite{Allahyarov:2007/a} are 
inferior to a short-range cut-off. 
\cite{Caillol:1995/b,Allahyarov:2008/a} This conclusion was backed up by 
analysis of $p(r)$, which shows that associated ions are in close 
proximity, while free ions show a very broad distribution of distances 
to the nearest ion of opposite charge.

To some extent, this feature is accounted for in theoretical treatments 
based on an ion-pair configurational integral, since it is the Boltzmann 
factor at short-range which makes the most significant contribution, at 
least at low temperatures where association is prevalent. The precise 
value of the upper limit in the integral is not important, as already 
noted long ago. \cite{Fowler:1956/a} The agreement between Bjerrum 
theory and simulation under the physical conditions studied here is 
excellent. More sophisticated treatments, such as those developed by 
Fisher and Levin, \cite{Fisher:1993/a,Levin:1996/a} are scarcely 
required; of course, near to the coexistence region, ion-ion and 
ion-dipole interactions become of paramount importance. Additionally, 
the analysis of experimental dielectric and conductivity data at low 
temperature and moderate concentrations requires a detailed account of 
ion triples and higher (charged) clusters; \cite{Robinson:1959/a} such 
phenomena as conductivity minima in the vicinity of the coexistence 
region can be handled using suitable extensions of the DH theory that 
include the effects of association. 
\cite{Weingartner:2000/a,Schroer:2004/a}

A comparison of the Bjerrum theory and the Debye-H{\"u}ckel theory shows 
that there is a significant region of the phase diagram where the former 
predicts strong ion association (in agreement with simulations) but the 
latter indicates only weak ion-ion correlations. This implies that 
linearised Poisson-Boltzmann theories, which assume weak ion-ion 
correlations, should only be applied under conditions where ion 
association is actually known to be insignificant. Those conditions are 
identified in this work. Of course, these restrictions should also be 
observed in the application of the Derjaguin-Landau-Verwey-Overbeek 
theory because it incorporates Debye screening of colloidal interactions 
under the assumption of weak ion-ion correlations, at least for the case 
of spherical colloidal particles.

The Bjerrum theory was compared with experimental data for 
tetraalkylammonium salts in low-polarity solvents, and the agreement was 
found to be excellent. Fitting the association constants allowed a 
mapping between real systems and the restricted primitive model. 
Simulation, theory, and experiment have therefore been compared at very 
low concentrations and low temperatures. One of the primary motivations 
for this study was to understand the association of ions in low-polarity 
solvents, with a view to getting a clear picture of the nature of 
screening between colloidal particles suspended in such solutions. The 
effects of ion association on the appropriate screening lengths have 
been quantified, and the results are in complete accord with experiment. 
It is hoped that the results presented herein can be used in subsequent 
treatments of charged-colloid interactions in low-polarity media.
	
\acknowledgments

The authors thank Dr Mirjam E. Leunissen and Dr Andrew D. Hollingsworth 
for communicating the results of Ref.~\onlinecite{Hollingsworth:2010/a} 
prior to publication, and Dr Leunissen and the anonymous referees for 
useful comments and suggestions. M.\ D.\ and C.\ V.\ acknowledge a 
NWO-VICI grant which supported the early stages of this work. C.\ V.\ 
acknowledges financial support from EPSRC grant EP/E030173, and from the 
European Union through an Individual Inter-European Marie Curie 
Fellowship. This work has made use of the resources provided by the 
Edinburgh Compute and Data Facility (ECDF). The ECDF is partially 
supported by the eDIKT initiative.

%%% Figure captions %%%

% \clearpage

\onecolumngrid

\section*{Figure captions}

\begin{list}{}{\leftmargin 1.5cm \labelwidth 1.5cm \labelsep 0.25cm}

\item[\bf Fig. 1]{The degree of association $\alpha$ along isotherms. 
The curves are the theoretical predictions of Eq.~(\ref{eqn:alpha}) 
using $K$ computed with a cut-off $r_{\rm c}=\lambda_{\rm B}/2$ (solid 
lines), and using the asymptotic result in Eq.~(\ref{eqn:kasymp}) for 
$K$ (dashed lines). The points are the simulation results computed using 
various criteria: (a) distance criterion with $r_{\rm c}=\lambda_{\rm 
B}$; (b) distance criterion with $r_{\rm c}=\lambda_{\rm B}/2$; (c) 
distance criterion with $r_{\rm c}=2\sigma$; (d) fitting 
Eq.~(\ref{eqn:prd}) to $p(r)$.}

\item[\bf Fig. 2]{Nearest-neighbour distribution function $p(r)$ at 
$T^{*}=0.05$. The simulation results are from simulations with 
$\rho^{*}=1.05 \times 10^{-6}$ (open symbols) and $\rho^{*}=1.00 \times 
10^{-4}$ (filled symbols); the thick curves are fits using the 
dissociated-ion result in Eq.~(\ref{eqn:prd}) over the range $r>r_{0}$, 
where $r_{0}$ is the maximum in $p(r)$.}

\item[\bf Fig. 3]{Cation-anion configurational integral, $K$, as a 
function of temperature $T^{*}$: Eq.~(\ref{eqn:k}) with $r_{\rm 
c}=\lambda_{\rm B}$ (solid line); Eq.~(\ref{eqn:k}) with $r_{\rm 
c}=\lambda_{\rm B}/2$ (dashed line) (this is almost indistinguishable 
from the former curve, in this temperature range); Eq.(\ref{eqn:kasymp}) 
(dot-dashed line).}

\item[\bf Fig. 4]{Phase diagram of the RPM showing the vapour-liquid 
coexistence region (lower right) from simulations, \cite{Romero:2002/a} 
and the boundary between dissociated ($\alpha < \frac{1}{2}$) and 
associated ($\alpha > \frac{1}{2}$) regimes as predicted from 
Eq.~(\ref{eqn:bjlocus}) with three different evaluations of $K$: 
Eq.~(\ref{eqn:k}) with $r_{\rm c}=\lambda_{\rm B}$ (black solid line) 
and $r_{\rm c}=\lambda_{\rm B}/2$ (red dotted line); 
Eq.~(\ref{eqn:kasymp}) (green short-dashed line). Also shown is the 
locus of maxima in the constant-volume heat capacity $C_{V}$ from a 
simple two-particle theory \cite{PJC:1999/a} (blue long-dashed line) and 
the boundary between ideal and strongly correlated regimes predicted by 
DH theory from Eq.~(\ref{eqn:dhlocus}) (black dot-dashed line). Real 
units are shown for a monovalent salt with ionic diameter 
$\sigma=4~\mbox{\AA}$ at a temperature $T=298.15~\mbox{K}$.}

\item[\bf Fig. 5]{Degree of association $\alpha$ against $K\rho = 
2K_{\rm M}c_{\rm s}/c^{\minuso}$ for some tetraalkylammonium salts in 
various solvents. References for the experimental data are given in 
square brackets. The solid line is Eq.~(\ref{eqn:alpha}). 
($\mbox{Bu}=$butyl, $\mbox{Pr}=$propyl, $\mbox{Pi}=$picrate.)}

\item[\bf Fig. 6]{The Debye length $\kappa_{\rm D}^{-1}$, and the 
effective screening length $\kappa^{-1}$ from Eq.~(\ref{eqn:kappaeff}), 
for a solution with $\sigma=4~\mbox{\AA}$, $\epsilon=7$, and 
$T=298.15~\mbox{K}$; this set of parameters corresponds to a reduced RPM 
temperature of $T^{*}=0.05$.}

\end{list}


\begin{thebibliography}{10}

\bibitem{Verwey:1948/a}
E.~J.~W. Verwey and J.~T.~G. Overbeek, {\em Theory of the stability of
  lyophobic colloids}, Elsevier, Amsterdam, 1948.

\bibitem{Yethiraj:2003/a}
A.~Yethiraj and A.~\protect{van Blaaderen}, {\em Nature}, 2003, {\bf 421},
  513--517.

\bibitem{Royall:2003/a}
C.~P. Royall, M.~E. Leunissen, and A.~\protect{van Blaaderen}, {\em J. Phys.:
  Condens. Matter}, 2003, {\bf 15}, S3581--S3596.

\bibitem{Leunissen:2005/a}
M.~E. Leunissen, C.~G. Christova, A.-P. Hynninen, C.~P. Royall, A.~I. Campbell,
  A.~Imhof, M.~Dijkstra, R.~\protect{van Roij}, and A.~\protect{van Blaaderen},
  {\em Nature}, 2005, {\bf 437}, 235--240.

\bibitem{Campbell:2005/a}
A.~I. Campbell, V.~J. Anderson, J.~S. \protect{van Duijneveldt}, and
  P.~Bartlett, {\em Phys. Rev. Lett.}, 2005, {\bf 94}, 208301.

\bibitem{Shevchenko:2006/a}
E.~V. Shevchenko, D.~V. Talapin, N.~A. Kotov, S.~O'Brien, and C.~B. Murray,
  {\em Nature}, 2006, {\bf 439}, 55--59.

\bibitem{Lu:2006/a}
P.~J. Lu, J.~C. Conrad, H.~M. Wyss, A.~B. Schofield, and D.~A. Weitz, {\em
  Phys. Rev. Lett.}, 2006, {\bf 96}, 028306.

\bibitem{Leunissen:2007/a}
M.~E. Leunissen, A.~\protect{van Blaaderen}, A.~D. Hollingsworth, M.~T.
  Sullivan, and P.~M. Chaikin, {\em Proc. Natl. Acad. Sci. (USA)}, 2007, {\bf
  104}, 2585--2590.

\bibitem{Leunissen:2007/b}
M.~E. Leunissen, J.~Zwanikken, R.~\protect{van Roij}, P.~M. Chaikin, and
  A.~\protect{van Blaaderen}, {\em Phys. Chem. Chem. Phys.}, 2007, {\bf 9},
  6405--6414.

\bibitem{Leunissen:2007/c}
M.~E. Leunissen {\em Manipulating colloids with charges and electric fields}
  PhD thesis, Utrecht University, 2007.

\bibitem{Zwanikken:2007/a}
J.~Zwanikken and R.~van Roij, {\em Phys. Rev. Lett.}, 2007, {\bf 99}, 178301.

\bibitem{Allahyarov:2007/a}
E.~Allahyarov, E.~Zaccarelli, F.~Sciortino, P.~Tartaglia, and H.~L{\"o}wen,
  {\em Europhys. Lett.}, 2007, {\bf 78}, 38002.

\bibitem{Allahyarov:2008/a}
E.~Allahyarov, E.~Zaccarelli, F.~Sciortino, P.~Tartaglia, and H.~L{\"o}wen,
  {\em Europhys. Lett.}, 2008, {\bf 81}, 59901.

\bibitem{Zwanikken:2009/a}
J.~Zwanikken and R.~van Roij, {\em J. Phys.: Condens. Matter}, 2009, {\bf 21},
  424102.

\bibitem{Robinson:1959/a}
R.~A. Robinson and R.~H. Stokes, {\em Electrolyte solutions}, Butterworths,
  London, 2nd ed., 1959.

\bibitem{Bjerrum:1926/a}
N.~Bjerrum, {\em Kgl.\ Dan.\ Vidensk.\ Selsk.\ Mat.-fys.\ Medd.}, 1926, {\bf
  7}, 1--48.

\bibitem{Pitzer:1984/a}
K.~S. Pitzer, {\em J. Phys. Chem.}, 1984, {\bf 88}, 2689--2697.

\bibitem{Pitzer:1990/a}
K.~S. Pitzer, {\em Acc. Chem. Res.}, 1990, {\bf 23}, 333--338.

\bibitem{Fisher:1993/a}
M.~E. Fisher and Y.~Levin, {\em Phys.\ Rev.\ Lett.}, 1993, {\bf 71},
  3826--3829.

\bibitem{Levin:1996/a}
Y.~Levin and M.~E. Fisher, {\em Physica A}, 1996, {\bf 225}, 164--220.

\bibitem{PJC:2010/b}
C.~Valeriani, P.~J. Camp, J.~W. Zwanikken, R.~van Roij, and M.~Dijkstra, {\em
  J. Phys.: Condens. Matter}, 2010, {\bf 22}, 104122.

\bibitem{Romero:2002/a}
J.~M. Romero-Enrique, L.~F. Rull, and A.~Z. Panagiotopoulos, {\em Phys. Rev.
  E}, 2002, {\bf 66}, 041204.

\bibitem{Vega:2003/a}
C.~Vega, J.~L.~F. Abascal, C.~McBride, and F.~Bresme, {\em J. Chem. Phys.},
  2003, {\bf 119}, 964--971.

\bibitem{Caillol:2002/a}
J.-M. Caillol, D.~Levesque, and J.-J. Weis, {\em J. Chem.\ Phys.}, 2002, {\bf
  116}, 10794--10800.

\bibitem{Luijten:2002/b}
E.~Luijten, M.~E. Fisher, and A.~Z. Panagiotopoulos, {\em Phys.\ Rev.\ Lett.},
  2002, {\bf 88}, 185701.

\bibitem{Valleau:1980/a}
J.~P. Valleau, L.~K. Cohen, and D.~N. Card, {\em J. Chem. Phys.}, 1980, {\bf
  72}, 5942.

\bibitem{Gillan:1983/a}
M.~J. Gillan, {\em Molec. Phys.}, 1983, {\bf 49}, 421--442.

\bibitem{Caillol:1995/a}
J.~M. Caillol, {\em J. Chem.\ Phys.}, 1995, {\bf 102}, 5471--5479.

\bibitem{Bresme:1995/a}
F.~Bresme, E.~Lomba, J.-J. Weis, and J.~L.~F. Abascal, {\em Phys. Rev. E},
  1995, {\bf 51}, 289--296.

\bibitem{PJC:1999/a}
P.~J. Camp and G.~N. Patey, {\em Phys.\ Rev.\ E}, 1999, {\bf 60}, 1063--1066.

\bibitem{PJC:1999/c}
P.~J. Camp and G.~N. Patey, {\em J. Chem.\ Phys.}, 1999, {\bf 111}, 9000--9008.

\bibitem{Shelley:1995/a}
J.~C. Shelley and G.~N. Patey, {\em J. Chem.\ Phys.}, 1995, {\bf 103},
  8299--8301.

\bibitem{PJC:2003/d}
C.~D. Daub, G.~N. Patey, and P.~J. Camp, {\em J. Chem. Phys.}, 2003, {\bf 119},
  7952--7956.

\bibitem{PJC:2007/b}
G.~Ganzenm{\"u}ller and P.~J. Camp, {\em J. Chem. Phys.}, 2007, {\bf 126},
  191104.

\bibitem{Hollingsworth:2010/a}
A.~D. Hollingsworth, M.~E. Leunissen, A.~Yethiraj, A.~\protect{van Blaaderen},
  and P.~Chaikin, unpublished work.

\bibitem{Fowler:1956/a}
R.~Fowler and E.~A. Guggenheim, {\em Statistical Thermodynamics}, Cambridge
  University Press, Cambridge, 1956.

\bibitem{Jordan:1973/a}
P.~C. Jordan, {\em Mol.\ Phys.}, 1973, {\bf 25}, 961--973.

\bibitem{Ebeling:1968/a}
W.~Ebeling, {\em Z. Phys.\ Chem.}, 1968, {\bf 238}, 400--402.

\bibitem{Orkoulas:1994/a}
G.~Orkoulas and A.~Z. Panagiotopoulos, {\em J. Chem.\ Phys.}, 1994, {\bf 101},
  1452--1459.

\bibitem{Allen:1987/a}
M.~P. Allen and D.~J. Tildesley, {\em Computer simulation of liquids},
  Clarendon Press, Oxford, 1987.

\bibitem{Caillol:1995/b}
J.-M. Caillol and J.-J. Weis, {\em J. Chem.\ Phys.}, 1995, {\bf 102},
  7610--7621.

\bibitem{Debye:1923/a}
P.~Debye and E.~H{\"{u}}ckel, {\em Phys.\ Z.}, 1923, {\bf 24}, 185--206.

\bibitem{McQuarrie:1976/a}
D.~A. McQuarrie, {\em Statistical mechanics}, Harper-Collins, New York, 1976.

\bibitem{Lindback:1980/a}
T.~Lindb{\"a}ck and P.~Beronius, {\em Acta Chem. Scand. A}, 1980, {\bf 34},
  709--715.

\bibitem{Roy:2009/a}
M.~N. Roy, P.~K. Roy, R.~S. Sah, P.~Pradhan, and B.~Sinha, {\em J. Chem. Eng.
  Data}, 2009, {\bf 54}, 2429--2435.

\bibitem{Atkins:2010/a}
P.~Atkins and J.~\protect{de Paula}, {\em Atkins' Physical Chemistry}, Oxford
  University Press, Oxford, 9th ed., 2010.

\bibitem{Weingartner:2000/a}
H.~Weing{\"{a}}rtner, V.~C. Weiss, and W.~Schr{\"{o}}er, {\em J. Chem. Phys.},
  2000, {\bf 113}, 762--770.

\bibitem{Schroer:2004/a}
W.~Schr{\"{o}}er and H.~Weing{\"a}rtner, {\em Pure Appl. Chem.}, 2004, {\bf
  76}, 19--27.

\end{thebibliography}
\end{document}